\documentclass[twocolumn,twoside,slac]{revtex4}
\usepackage{graphicx}
\usepackage{fancyhdr}
\begin{document}

\title{Commissioning the CDF Offline Software}


\author{Elizabeth Sexton-Kennedy, Pasha Murat}
\affiliation{FNAL, CD/CDF, Batavia, IL 60510, USA}

\begin{abstract}

CDF II is one of the two large collider experiments at Fermilab's Tevatron.
Over the past two years we have commissioned the offline computing system.  A
task that has involved bringing up hundreds of computers and millions of lines
of C++ software.  This paper reports on this experience, concentrating on the
software aspects of the project.  We will highlight some of the successes as
well as describe some of the work still to do.

\end{abstract}

\maketitle

\thispagestyle{fancy}


\section{SCOPE AND SCALE OF THE OFFLINE SOFTWARE PROJECT}

	The CDF offline involves hundreds of collaborators from 56 different
institutions from all over the world.  While not ever collaborator has directly
contributed software, all must use the software and help debug it.  The current
code set consists of tens of millions of mostly C++ code, organized into 294
packages.  These packages are further organized into major reconstruction,
simulation, and physics categories: Tracking, Calorimetry, Muon, Time of Flight,
Luminosity counters, Top, Tau, Electroweak, B physics, and infrastructure.
	At CDF the offline includes all of the software and computing need for
the real time software trigger (the Level3 trigger), the online monitoring of
the detector, the primary reconstruction, the detector simulation, physics group
analysis software, tools such as the detector event display and software used to
measure the performance of the above, such as tracking efficiency and purity.

\section{HISTORY OF THE PROJECT}

\subsection{Manpower}
	In 1996 when work on the run 2 offline started, very few people at CDF
knew C++.  The advantages of using a language that supported memory management
was considered large enough to out way the cost of learning C++.  We claimed
that we would start by writing all of the infrastructure code in C++ but still
make provisions for the physics code to be written in f77.  People involved
with the run 1 experiment wanted to reuse as much of the run 1 code as
possible, so the code was wrapped into C callable routines and fed it's input
data via the C++ infrastructure.  Output was in the form of arrays that were
also managed by the infrastructure.  As new people entered the project there was
a desire to replace this code with algorithms written in C++ so that it would
be maintainable.  This happened for every subsystem until everything was C++.
The next generation of maintainers after this second round has not felt it
necessary to rewrite everything.  We now have people entering the project with
prior C++ experience from other experiments.
	During the development of the code we never had enough people to work
on all of the tasks we wanted done.  Strong leadership to prioritize during
these early times was essential.  At any one time we had about 6 to 10 highly
productive developers, one per subsystem.

\subsection{Releases and Transitions}
The release schedule reflects the pace of development and managements decisions
about how to best serve our users.  Our users were the detector commissioning
physicists, and farm and data handling hardware developers. 
During the first year that we started making major releases, 1999, there were 5
releases reflecting the difficulty of these first integrations.  In 2000 and 
2001 there were about 10 each year.  This was our time
of extreme programing.  Many of the major rewrites occurred at this time,
including a change from a Fortran array based EDM (Event Data Model) to a root
based EDM.  At the same time the experiment was commissioning the detector and
these customers required that we keep the code working for their use.  For this
reason many subsystem developers were maintaining two versions of the code for
their system.  One of the advantages of this situation was that both versions
could be run to see if they resulted in the same answer.  Many bugs were
discovered while investigating differing answers.  Some were in the C++ but
some were found in the run1 Fortran code as well.  In 2002 there were about 5
releases.  This reflected the need for stability in preparation for presenting
physics results at the early 2003 winter conferences.

\subsection{Collaboration with Others}
Throughout the history of the project CDF was greatly aided by using the work
of others outside of CDF.  BaBar donated their framework, EvtGen and ideas.
The support of the root team especially Phillippe Canal was essential.  Many of
the Zoom/CLHEP classes we use were developed by those groups at our request.
Many of the classes in existence before CDF started developing C++ software were 
greatly improved in performance through collaboration with the supporters of CLHEP.
Unfortunately our compiler vendor, KAI was also a collaborator in that we had to
report bugs to them.  Despite this, the decision to use KAI was not a bad one.
It allowed us to write C++ standard code much earlier then we would otherwise
have been able to.  This has served us well in transitioning back to g++ now that
KAI is becoming unsupported.

\section{DEVELOPMENT OF STABLE OPERATIONS}
The most important step in creating stability and robustness in the software
system was the development of rules and procedures.  There are rules about how
releases are put together and what can be integrated at different phases of the
cycle.  There are well defined procedures for validation and regression
testing of all new releases.  These are documented on the web and the manpower
for doing these tasks comes from the collaboration as part of an offline shift.
A shifter is given recipes for running purify, debuggers and software
management tools as well as instructions for running the tests.  Monitoring the 
running of automated systems like the reconstruction farms is also part of the
job.  Use of a bug tracking tool has also greatly aided in finding problems and
documenting solutions for out users.  The shifter can try to answer questions
that are sent to the list, or forward them to the relevant system expert.

\section{LESSONS LEARNED AND SURPRISES}
Here is a list of some of the things that were surprising and were learned over
the course of the project: 
\begin{itemize}
\item You never really understand a problem until it is solved once.  The
strategy of studying and recoding run 1 code in C++ for run 2 served us well in
many areas of the project.
\item Performance has not been a problem, wasteful copying was eliminated early
in the development of the code and there are still gains that we can make.
Choosing efficient algorithms gains more performance the hand optimizing the
code.
\item Keep the system clean in terms of physical design and organization.
Physics analysis codes will follow the patterns of the reconstruction and be
more generally usable if the reconstruction is kept clean.  
\item A code browser is important for both the developers and users of the
software system.
\item Beware of code generation, it can produce code bloat if not done carefully.
\item Memory leaks are supposed to be the biggest problem in large C++ systems,
however for us uninitialized variables has been a bigger problem.  Memory leaks
can be traced with standard tools.  Uninitialized memory reports from these same
tools are too numerous to be useful.  Many of these reports are completely
harmless and finding the bad ones buried underneath is very hard.  We have had
releases in which the farms operation of the reconstruction was tested to be 1
crash in a million events.  The same exact code when recompiled with a shorter
name for the file system it lived on crashed in less then a thousand events
processed.  This was traced to an error in using uninitialized memory.
\item Start commissioning as early as possible.  Mock data challenges didn't
completely prepare us for the turn on of real data. Count on having to change
things once you have really customers with real needs.
\item Many people will leave the project one year after the first physics data 
arrives.  If faced with the choice of two solutions to a problem always choose
the most maintainable one.  Insure that there is sufficient overlap between the
original developers of the code and the junior people who will take over the
maintenance of it.
\end{itemize}

\section{THE CDF OFFLINE SOFTWARE IS A SUCCESS}
The CDF offline software system was flexible and serviceable.  It was able to
help commission the detector even while it was undergoing large transitions to
improved software.  It helped take the data, reconstruct it, and perform the
physics analysis on it in a timely manner.

\begin{acknowledgments}
The authors wish to thank all of the people who worked on the CDF
offline software.
\end{acknowledgments}

%\begin{thebibliography}{9}   % Use for  1-9  references
%\begin{thebibliography}{99} % Use for 10-99 references

%\bibitem{accelconf-ref}
%http://www.cern.ch/accelconf

%\bibitem{exampl-ref}
%A.N. Other, ``A Very Interesting Paper'', EPAC'96, Sitges, June
%1996.

%\bibitem{templates-ref}
%http://www.cern.ch/accelconf/templates.html

%\end{thebibliography}

\end{document}